\documentclass{cmmp}
\begin{document}
\chapter*{The quantum physics of chronology protection}
\vskip -2cm
\centerline{Matt Visser}
\vskip 2mm
\centerline{\small\it 
Physics Department, Washington University, 
Saint~Louis, Missouri 63130-4899, USA.}
\vskip 2mm
\centerline{{\small 2 April 2002; Additional references 17 April 2002; \LaTeX-ed \today}}
\vskip 5mm
{\bf Abstract:}

This is a brief survey of the current status of Stephen Hawking's
``chronology protection conjecture''. That is: ``Why does nature abhor
a time machine?'' I'll discuss a few examples of spacetimes containing
``time machines'' (closed causal curves), the sorts of peculiarities
that arise, and the reactions of the physics community.  While
pointing out other possibilities, this article concentrates on the
possibility of ``chronology protection''. As Stephen puts it:
\begin{quotation}
\emph{It seems that there is a Chronology Protection Agency which
prevents the appearance of closed timelike curves and so makes the
universe safe for historians.}
\end{quotation}

\vskip 1 cm
\noindent
To appear in: {\bf The future of theoretical physics and cosmology};\\ 
Proceedings of the conference held in honour of Stephen Hawking on the
occasion of his 60'th birthday. (Cambridge, 7--10 January 2002.)

\vspace*{5mm}
\noindent

\vfill
\hrule
\bigskip
\centerline{\underline{E-mail:} {\sf visser@kiwi.wustl.edu}}
\bigskip
\centerline{\underline{Homepage:} 
{\sf http://www.physics.wustl.edu/\~{}visser}}
\bigskip
\centerline{\underline{Archive:} {\sf   gr-qc/0204022}}
\bigskip
\noindent
{\small \it Permanent address after 1 July 2002:
School of Mathematics and Computer Science, Victoria University, 
PO Box 600, Wellington, New Zealand. {\sf matt.visser@mcs.vuw.ac.nz} }
\bigskip
\hrule
\clearpage

\newcommand{\be}{\begin{equation}}
\newcommand{\ee}{\end{equation}}
\let\Rightarrow=\implies
\def\Godel{G\"odel}
\def\d{{\mathrm{d}}}
\newtheorem{theorem}{Theorem}
\newtheorem{corollary}{Corollary}
\setcounter{chapter}{0}
\chapter*{The quantum physics of chronology protection}

Simply put, chronology protection is the assertion that nature abhors
a time machine. In the words of Stephen Hawking~\cite{Stephen-CPC}:
\begin{quotation}
\emph{It seems that there is a Chronology Protection Agency which
prevents the appearance of closed timelike curves and so makes the
universe safe for historians.}
\end{quotation}
The idea of chronology protection gained considerable currency during
the 1990's when it became clear that traversable wormholes, which are
not too objectionable in their own
right~\cite{Morris-Thorne,surgery,examples}, seem almost generically
to lead to the creation of time
machines~\cite{MTY,Thorne,debate,Lorentzian}. The key word here is
``seem''. There are by now many technical discussions available in the
literature (well over 200 articles), and in the present chapter I will
simply give a pedagogical and discursive overview, while adding an
extensive bibliography for those interested in the technical details.
First: a matter of language, for all practical purposes the phrases
``time machine'' and ``closed timelike curve'' (or the closely related
``closed null curve'') can be used interchangeably.

\section*{Why is chronology protection even an issue?}

Before embarking on a discussion of chronology and how it is believed
to be protected~\cite{Stephen-CPC,Thorne,debate}, it is useful to
first ask why chronology even needs to be protected. In Newtonian
physics, and even in special relativity or flat-space quantum field
theory, notions of chronology and causality are so fundamental that
they are simply built into the theory ab initio. Violation of normal
chronology (for instance, an effect preceding its cause) is so
objectionable an occurrence that any such theory would immediately be
rejected as unphysical.

Unfortunately, in general relativity one cannot simply \emph{assert\/}
that chron\-ology is preserved, and causality respected, without doing
considerable additional work. The essence of the problem lies in the
fact that the Einstein equations of general relativity are local
equations, relating some aspects of the spacetime curvature at a point
to the presence of stress-energy at that point. Additionally, one also
has local chronology protection, inherited from the fact that the
spacetime is locally Minkowski (the Einstein Equivalence Principle),
and so ``in the small'' general relativity respects all of the
causality constraints of special relativity.

What general relativity does \emph{not} do is to provide any natural
way of imposing \emph{global} constraints on the spacetime ---
certainly the Einstein equations provide no such nonlocal
constraint. In cosmology this leads to the observation that the global
topology of space is not constrained by the Einstein equations;
spatial topology is an independent discrete variable that has to be
decided by observation. (And this requires additional data over and
above whatever is needed to decide the familiar $k=+1$, $k=0$, or
$k=-1$ question of the Friedmann--Robertson--Walker
cosmologies~\cite{Cosmic-Topology}.) Similarly, global temporal
topology is not constrained by the Einstein equations themselves, and
additional physical principles need to be brought into play to somehow
deal with the possibility of nontrivial-temporal topology.

Without imposing additional principles along these lines general
relativity is completely infested with time machines (in the sense of
closed causal curves). Perhaps the earliest examples of this pathology
are the Van~Stockum spacetimes~\cite{van-Stockum}, but the example
that has attracted considerably more attention is Kurt \Godel's
peculiar cosmological solution~\cite{Godel}. These spacetimes are
exact solutions of the Einstein equations, with sources that (at least
locally) look physically reasonable, which nevertheless possess
serious global pathologies. If it were only a matter of dealing with
these two particular examples, physicists would not be too worried ---
but similar behaviour occurs in many other geometries, for instance,
deep inside the Kerr solution.  A complete list of standard but
temporally ill-behaved spacetimes is tedious to assemble, but at a
minimum should include:
\begin{enumerate}
\item
\Godel's cosmology~\cite{Godel};
\item
Van Stockum spacetimes~\cite{van-Stockum}/ 
Tipler cylinders~\cite{Tipler-c}/
\\ 
longitudinally spinning cosmic strings~\cite{Lorentzian};
\item
Kerr and Kerr--Newman geometries~\cite{Hawking-Ellis};
\item
Gott's time machines~\cite{Gott};
\item
Wheeler wormholes (spacetime foam)~\cite{Wheeler,Wheeler2,Geometrodynamics};
\item
Morris--Thorne traversable wormholes~\cite{Morris-Thorne,MTY};
\item
Alcubierre ``warp drive'' spacetimes~\cite{Warp}.
\end{enumerate}
The Wheeler wormholes are included based on theorems that localized
topology change implies either causal pathology or naked
singularities; either possibility is
objectionable~\cite{Tipler-s,Tipler-s2,Tipler-s3}.  The Morris--Thorne
traversable wormholes are included based on the observation that
apparently trivial manipulations of these otherwise not too
objectional geometries seem to almost generically lead to the
development of closed timelike curves and the destruction of normal
chronology~\cite{MTY,Lorentzian}. For the ``warp drive'' spacetimes
manipulations similar to those performed for traversable wormhole
spacetimes seem to lead inevitably to time travel. (Once one has
effective faster-than-light travel, whether via wormholes or
warpdrives, the twin pseudo-paradox of special relativity is converted
into a true paradox, in the sense of engendering various time travel
paradoxes.)

Now in each of these particular cases you can at a pinch find
\emph{some} excuse for not being too concerned, but it's a different
excuse in each case. The matter sources for the {\Godel} solution are
quite reasonable, but the observed universe simply does not have those
features. The Van~Stockum time machines and their brethren require
infinitely long cylindrical assemblages of matter rotating at
improbable rates.  Gott's time machines have pathological and
non-physical global behaviour~\cite{Deser1,Deser2}.  The Kerr and
Kerr--Newman pathologies are safely hidden behind the Cauchy
horizon~\cite{Hawking-Ellis}, where one should not trust naive notions
of maximal analytic extension. (The inner event horizon is classically
unstable.) The Wheeler wormholes (spacetime foam) have never been
detected, and at least some authors now argue against the very
existence of spacetime foam. The energy condition violations implicit
in traversable wormholes and warp drive spacetimes do not seem to be
qualitatively insurmountable problems~\cite{Morris-Thorne,Lorentzian},
but do certainly give one pause~\cite{Cosmo99}. This multiplicity of
different excuses does rather make one worry just a little that
something deeper is going on; and that there is a more general
underlying theme to these issues of (global) chronology protection.

\section*{Paradoxes and responses.}

Most physicists view time travel as being problematic, if not
downright repugnant.  There are two broad classes of paradox generated
by the possibility of time travel, either one of which is
disturbing:
\begin{enumerate}
\item 
Grandfather paradoxes: Caused by attempts to ``change the past'', and
so modify the conditions that lead to the very existence of the entity
that is trying to ``modify the timestream''.
\item
Bootstrap paradoxes: Where an effect is its own cause.
\end{enumerate}
Faced with the {\em a priori} plethora of geometries containing closed
timelike curves, with the risk of these two classes of logical paradox
arising, the physics community has developed at least four distinct
reactions~\cite{Lorentzian}:
\begin{enumerate}
\item
Make radical alterations to our worldview to incorporate at least some
versions of chronology violation and ``time travel''. (The ``radical
re-write'' conjecture.)  One version of the radical re-write
conjecture uses non-Hausdorff manifolds to describe ``train track''
geometries where the same present has two or more futures (or two or
more pasts). A slightly different version uses the ``many worlds''
interpretation of quantum mechanics to effectively permit switching
from one history to another~\cite{Deutsch}. More radically one can
even contemplate multiple coexisting versions of the ``present''.
\item
Permit constrained versions of closed timelike curves --- supplemented
with a consistency condition that essentially prevents any alteration
of the past. (This is the essence of the Novikov consistency
conditions~\cite{Novikov,Novikov2,Novikov3}.)
The consistency conditions are sometimes summarized as ``you can't
change recorded history''~\cite{LSH}. The central idea is that there is
a single unique timeline so that even in the presence of closed
timelike curves there are constraints on the possibilities that can
occur. In idealized circumstances these consistency constraints can be
derived from a least action principle. More complicated situations
seem to run afoul of the notion of ``free will'', though there is
considerable doubt as to the meaning of ``free will'' in the presence
of time travel~\cite{Zombies}.
\item
Appeal to quantum physics to intervene and provide a universal
mechanism for preventing the occurrence of closed timelike curves.
This, in a nutshell, is Stephen Hawking's ``chronology protection''
option, the central theme of this chapter, which we shall develop in
considerable detail below.
\item
Agree to not think about these issues until the experimental evidence
becomes overwhelming. (The ``boring physics'' conjecture.)  After all,
what is the current experimental evidence? Assume global hyperbolicity
and cosmic censorship and be done with it. If, for instance, one takes
canonical gravity seriously as a fundamental theory then there exists
at least one universal foliation by complete spacelike
hypersurfaces. This automatically forbids closed timelike curves at
the kinematical level, before dynamics (classical or quantum) comes
into play. However, it should be noted that canonical gravity
interpreted in this strict sense has severe difficulties (for
instance, in dealing with maximal analytic extensions of the Kerr
spacetime).
\end{enumerate}
Originally it was hoped that it would be possible to decide between
these options based on classical or at worst semiclassical physics ---
however it it now becoming increasingly clear that the ultimate
resolution of the chronology protection issue will involve deep
issues of principle at the very foundations of the full theory of
quantum gravity.

\section*{Elements of chronology protection.}

Chronology protection is at one level an attempt at ``having one's
cake and eating it too'' --- this in the sense that it provides a
framework sufficiently general to permit interesting and nontrivial
topologies and geometries, but seeks to keep the unpleasant side
effects under control. Chronology protection deals with the localized
production and destruction of closed timelike curves; the very essence
of what we might like to think of as ``creating'' a time machine.

(Cosmological time machines, in the sense of {\Godel}, are best viewed as
an example of the GIGO principle; garbage in, garbage out. Just
because one has a formal solution to a set of differential equations
does not mean there is any physical validity to the resulting
spacetime. A differential equation without boundary conditions/
initial conditions has little predictive power, and it is very easy to
generate ill-posed problems. Cosmological time machines are by
definition intrinsically and equally sick everywhere in the
spacetime.)

In the case of a localized production of closed timelike curves the
situation is more promising: the spacetime is then divided into
regions of normal causal behaviour and abnormal causal behaviour, with
the boundary that separates these regions referred to as the
``chronology horizon''. It is the behaviour of quantum physics at and
near this chronology horizon that provides the basis for chronology
protection.

Specifically, a point $x$ is part of the chronology violating region
if there is a closed causal curve (closed timelike curve) or closed
chronological curve (closed null/ timelike curve) passing through
$x$. The chronology horizon is then defined as the boundary of the
future of the chronology violating region. (That is, the boundary of
the region from which chronology violating physics is visible.) This
chronology horizon is by definition a special type of Cauchy
horizon. Under reasonably mild technical conditions Hawking has argued
that the chronology horizons appropriate to locally constructed time
machines should be compactly generated and contain a ``fountain'';
essentially the first closed null curve to come into existence as the
time machine is formed~\cite{Stephen-CPC}.

A classical photon placed on this fountain will circulate around the
fountain infinitely many times; in effectively zero ``elapsed''
time. On each circuit around the fountain there is generically a
nontrivial holonomy that changes the energy of the photon. For a past
chronology horizon, which expands as we move to the future (as defined
by someone outside the chronology violating region) this provides a
boost, a net increase in the photon energy for each circuit of the
fountain. The photon energy increases geometrically, reaching infinity
in effectively zero time~\cite{Stephen-CPC}.  On each circuit
\[
E \to e^h\; E \to e^{2h}\; E \;\dots; 
\qquad
h = - 2 \oint \Re(\epsilon) \;\d t,
\]
with the size of the energy boost being controlled by a loop
integration around the fountain involving the Newman--Penrose
parameter $\epsilon$.  (In simple situations involving wormholes this
holonomy is essentially the Doppler shift factor due to relative
motion of the wormhole mouths, but when phrased in terms of $\oint
\epsilon$ it can be generalized to arbitrary chronology horizons
possessing a fountain.) The source of this energy must ultimately be
the spacetime geometry responsible for the chronology horizon, and by
extension, the stress-energy used to warp spacetime and set up the
fountain in the first place. If we now let the photon (and the
gravitational field it generates) back-react on the spacetime, its
infinite energy will presumably alter the spacetime geometry beyond
all recognition.  

Unfortunately this is a classical argument, appropriate to a classical
point particle following a precisely defined null curve. Will quantum
physics amplify or ameliorate this effect? Real photons are
wave-packets with a certain transverse size, and generically the same
effect that leads to the energy being boosted leads to the wave-packet
being defocussed --- the geometry in a tubelike region surrounding the
fountain acts as a diverging lens~\cite{Stephen-CPC}.

With two competing effects, the question becomes which one wins? The
answer, ``it depends''. There are geometries for which the classical
defocussing effect overwhelms the boost effect, and the classical
stress tensor remains finite on the fountain. There are other
geometries for which the reverse holds true. But this certainly means
that classical effects do not provide a {\emph{universal}} mechanism
for eliminating all forms of closed causal curves. Thus the search for a
universal chronology protection mechanism must then (at the very
least) move to the semiclassical quantum realm.

\section*{Semiclassical arguments}

In semiclassical quantum gravity, one treats gravity as a classical
external field, but one quantizes everything else. So far, this is
just curved space quantum field theory. But then one additionally
demands that the Einstein equations hold for the quantum expectation
value of the stress-energy tensor:
\[
G_{\mu\nu} = 8\pi G_{Newton} \; \langle \psi |T_{\mu\nu} |\psi\rangle.
\]
Semiclassical quantum gravity seems [at first glance] to lead to a
universally true statement to the effect that the renormalized
expectation value of the stress-energy tensor blows up at the
chronology horizon. The idea is based on the fact that in curved
manifolds (modulo technical issues to be discussed below) the
two-point correlation function (Green function; a measure of the mean
square fluctuations) of any quantum field is of Hadamard form
\[
G(x,y) = \sum_\gamma {\Delta_\gamma(x,y)^{1/2}\over4\pi^2}
\left\{ {1\over\sigma_\gamma(x,y) }
+ \nu_\gamma(x,y)\;\ln|\sigma_\gamma(x,y)| 
+ \varpi_\gamma(x,y)
\right\}.
\]
Here the sum runs over the distinct geodesics from $x$ to $y$; the
quantity $\Delta_\gamma(x,y)$ denotes the Van~Vleck determinant
evaluated along the geodesic $\gamma$; the quantity
$\sigma_\gamma(x,y)$ denotes Synge's ``world function'' (half the
square of the geodesic distance from $x$ to $y$); and the two functions
$\nu_\gamma(x,y)$ and $\varpi_\gamma(x,y)$ are smooth with finite
limits as $y \to x$. Provided the Green function can be put into this
Hadamard form, the expectation value of the point split stress-energy
tensor can be defined by a construction of the type
\[
\langle T_{\mu\nu}(x,y,\gamma_0) \rangle = D_{\mu\nu}(x,y,\gamma_0) \; G(x,y).
\]
Here $\gamma_0$ denotes the trivial geodesic from $x$ to $y$ (which
collapses to a point as $y\to x$, this geodesic will be unique
provided $x$ and $y$ are sufficiently close to each other), while
$D_{\mu\nu}(x,y,\gamma_0)$ is a rather complicated second-order
differential operator built up out of covariant derivatives at $x$ and
$y$. The covariant derivatives at $y$ are parallel transported back to
$x$ along the geodesic $\gamma_0$ with the result that $\langle
T_{\mu\nu}(x,y,\gamma_0)\rangle$ is a tensor with respect to
coordinate changes at $x$, and a scalar with respect to coordinate
changes at $y$. One then defines the renormalized expectation value of
the stress-energy tensor by taking the limit $y\to x$ and discarding
the universal divergent piece which arises from the contribution of
the trivial geodesic to the Green function. In other words, the
renormalized Green function is defined by
\[
G(x,y)_R = \sum_{\gamma\neq\gamma_0} {\Delta_\gamma(x,y)^{1/2}\over4\pi^2}
\left\{ {1\over\sigma_\gamma(x,y) }
+ \nu_\gamma(x,y)\;\ln|\sigma_\gamma(x,y)| 
+ \varpi_\gamma(x,y)
\right\},
\]
and the renormalized stress energy by
\[
\langle T_{\mu\nu}(x) \rangle_R = \lim_{y\to x} D_{\mu\nu}(x,y,\gamma_0) \; G_R(x,y).
\]
Other methods of regularizing and renormalizing the stress-energy
could be used, the results will qualitatively remain the same. The net
result is that
\[
\langle T_{\mu\nu}(x) \rangle_R = 
\sum_{\gamma\neq\gamma_0} 
{\Delta_\gamma(x,x)^{1/2}\over\sigma_\gamma(x,x)^2} \; t_{\mu\nu}(x) 
+ \cdots
\]
Here $t_{\mu\nu}(x)$ is a dimensionless tensor built up out of the metric
and tangent vectors to the geodesic $\gamma$, while the $\cdots$
denote subdominant contributions. The key observation is that if any
of the non-trivial geodesics from $x$ to itself are null (invariant
length zero), then there is an additional infinity in the
stress-energy over and above the universal local contribution that was
removed by renormalization.  (For a slightly different way of doing
things, one could just as easily choose to work with the effective
action~\cite{action} instead of the stress-energy; the conclusions are
qualitatively similar.)

In general these self-intersecting null geodesics define the
$N$'th-polarized hypersurfaces, where $N$ is a winding number which
counts the number of times the geodesic passes through the tubular
region surrounding the fountain. These polarized hypersurfaces lie
inside the chronology horizon and typically approach it as
$N\to\infty$~\cite{Thorne,debate}. In particular, the fountain is a
nontrivial closed null geodesic, and this argument indicates that the
renormalized stress-energy tensor diverges at the fountain. But
infinite stress-energy implies, via the Einstein equations, infinite
curvature. The standard interpretation of this is (or rather, was)
that once back-reaction is taken into account the fountain (and
{\emph{ipso facto}}, the entire chronology horizon) is destroyed by
the (mean square) quantum fluctuations. (You do not need the
stress-energy to diverge everywhere on the chronology horizon; it is
sufficient if it diverges at the fountain.)

The fly in the ointment here is these same quantum fluctuations. On
the one hand the quantum fluctuations are responsible for the formal
infinity in the expectation value of the stress-energy at the
fountain, on the other hand: Does the back-reaction due to the
expectation value of the stress-energy tensor become large before the
quantum fluctuations in the metric completely invalidate the manifold
picture? (This very question led to a spirited debate between Stephen
Hawking and Kip Thorne~\cite{Thorne,debate}, with disagreement on how
to define the notion of ``closeness'' to the chronology horizon.)

It is now generally accepted that typically the back reaction becomes
large before metric fluctuations invalidate the manifold picture, but
that there are exceptional geometries where the back-reaction can be
kept arbitrarily small arbitrarily close to the chronology horizon. A
particular example of this phenomenon is if you take a ``ring
configuration'' of wormholes, where each individual wormhole is
nowhere near forming a chronology horizon, but the combination is just
on the verge of violating causality~\cite{Roman-ring}. Then there is a
closed spacelike geodesic which traverses the entire ring of wormholes
whose invariant length is becoming arbitrarily small; but because the
spacelike geodesic is traversing many wormhole mouths (each of which
acts as a defocussing lens) the Van~Vleck determinant can be made
arbitrarily small in compensation.

That is: adopt the length of the shortest closed spacelike geodesic as
a diagnostic for how close the spacetime is to forming a time
machine. Then no matter how close one is to violating chronology,
there are some geometries for which the renormalized stress-energy
tensor (and the quantum-induced back reaction) can be made arbitrarily
small.  In a similar vein there are a number of other special case
examples (for example, toy models based on variants of the Grant and
Misner spacetimes~\cite{Grant,Krasnikov,Automorphic,Misner}) for which
the renormalized stress-energy remains finite all the way up to the
chronology horizon.  The upshot of all this is that the search for a
universal chronology protection mechanism must (at the very least)
involve issues deeper and more fundamental than the size of the
quantum-induced back reaction.

\section*{The failure of semiclassical gravity}

The most mathematically precise and general statements known
concerning the nature of the pathology encountered at the chronology
horizon are encoded in the singularity theorems of Kay, Radzikowski,
and Wald~\cite{KRW}. In a highly technical article using micro-local
analysis they demonstrated:
\begin{theorem}
There are points on the chronology horizon where the two-point
function is not of Hadamard form.
\end{theorem}
Because there are points where the two-point function is not of
Hadamard form, the entire process of defining a renormalized
stress-energy tensor breaks down at those points.  That is:
\begin{corollary}
There are points on the chronology horizon where semiclassical
Einstein equations fail to hold.
\end{corollary}
Note that the semiclassical Einstein equations,
\[
G_{\mu\nu} = 8\pi G_{Newton} \; \langle T_{\mu\nu} \rangle_R,
\]
fail for a subtle reason; they fail simply because at some points the
RHS fails to exist, not necessarily because the RHS is infinite. Now
typically, based on the explicit calculations of the last section, the
renormalized stress-energy does blow up on parts of the chronology
horizon. The significant new feature of the Kay--Radzikowski--Wald
analysis is that even if the stress-energy remains finite as one
approaches the chronology horizon, there will be points on the
chronology horizon for which no meaningful limit exists. (For a
specific example, see~\cite{ill-defined}.)

The physical interpretation is that semiclassical quantum gravity
fails to hold (at some points) on the chronology horizon; a fact which
can be read in two possible ways:
\begin{enumerate}
\item
If you assume that semiclassical quantum gravity is the fundamental
theory (at best a minority opinion, and there are good very reasons
for believing that this is not the case), then by \emph{reductio ad
absurdum} the chronology horizon must fail to form. Chronology is
protected, essentially by \emph{fiat}.
\item
If you are willing to entertain the possibility that semiclassical
quantum gravity is not the whole story (the majority opinion), then it
follows from the above that issues of chronology protection cannot be
settled at the semiclassical level. Chronology protection must then be
settled (one way or another) at the level of a full theory of quantum
gravity.
\end{enumerate}

An attractive physical picture that captures the essence of the
situation is this: Sufficiently close to (but outside) the chronology
violating region there are extremely short self-intersecting spacelike
geodesics. The length of these geodesics can be used to develop an
observer independent measure of closeness to chronology
violation. Indeed let
\[
{\cal M}(\ell) = 
\left\{ x \;\Big|\; 
\exists \; \gamma \neq \gamma_0 \;:\; 
\sigma_\gamma(x,x) \leq {\ell^2\over2} \right\}
\]
Then ${\cal M}(0)$ is one way of characterizing the chronology
violating region, while ${\cal M}(L_{Planck})-{\cal M}(0)$ is an
invariantly defined region just outside the chronology violating
region which is covered by extremely short spacelike geodesics. In a
tubelike region along any one of these geodesics the metric can be put
in the form
\[
\d s^2 = \d l^2 + g^{(2+1)}_{ab}(l,x_\perp)\; \d x_\perp^a \;\d x_\perp^b,
\]
subject to the boundary condition
\[
g^{(2+1)}_{ab}(0,0_\perp) =  g^{(2+1)}_{ab}(\ell,0_\perp).
\]
If we now Fourier decompose the metric in this tubelike region the
boundary conditions imply that $p_\ell = n \hbar/\ell$.  For
$\ell<L_{Planck}$, high-momentum trans-Planckian modes $p_\ell > n
\hbar/L_{Planck} = n E_{Planck}/c$ are an unavoidable part of the
analysis. That is, close enough to the chronology violating region one
is intrinsically confronted with Planck scale physics; and the onset
of Planck-scale physics can be invariantly characterized by the length
of short but nontrivial spacelike geodesics. In particular the
relevant Planck scale physics includes Planck scale fluctuations in
the metric --- these fluctuations in the geometry of spacetime fuzz
out the manifold picture that is the essential backdrop of
semiclassical gravity. Thus quantum physics wins the day, and curved
space quantum field theory is simply not enough to complete the job.

Overall, this entire chain of development has led the community to a
conclusion diametrically opposed to the initial hopes of the early
1990's --- the hopes for a simple and universal classical or
semiclassical mechanism leading to chronology protection seem to be
dashed, and the relativity community is now faced with the daunting
prospect of understanding full quantum gravity just to place notions
of global causality on a firm footing.

\section*{Where we stand}

There is ample evidence that quantum field theory is a good
description of reality, and there is also ample evidence that general
relativity (Einstein gravity) is a good description of reality. From
the obvious statement that in our terrestrial environment gravity is
well described by classical general relativity, while condensed matter
physics is well described by quantum physics, it follows that
semiclassical quantum gravity (curved space quantum field theory with
the Einstein equations coupled to the quantum expectation value of the
stress-energy) is a more than adequate model over a wide range of
situations. (No-one seriously doubts the applicability of
semiclassical gravity to planets, stars, galaxies, or even to
cosmology itself once the universe has emerged from the Planck era.)

Nevertheless, there are apparently plausible situations in
semiclassical gravity that naively seem to lead to the onset of
causality violation; and attempts at protecting chronology inevitably
lead one back to considerations of full quantum gravity. The situation
is somewhat reminiscent of black hole physics where the infinite
redshift at the black hole horizon is often interpreted as a
microscope that could potentially open a window on the Planck
regime~\cite{'tHooft1,'tHooft2}.  Similarly, in discussing chronology
protection the region near the chronology horizon is subject to Planck
scale physics (believed to include Planck scale fluctuations in the
geometry) so that semiclassical gravity is not a ``reliable'' guide
near the chronology horizon~\cite{Reliable,Reliable2}.  This opens a
second window on Planck scale physics --- though the chances of
experimentally building a time machine (or getting close enough to
forming a chronology horizon to actually see what happens) must be
viewed as even somewhat less likely than the chances of experimentally
building a general relativity black hole. (Black hole analogues, such
as acoustic dumb holes, are another story~\cite{Unruh,Garay,Laval}.)

One possible response, given that we will inevitably have to face
full-fledged quantum gravity, is to take chronology protection as
being so basic a property that we should use it as a guide in
developing our theory of quantum gravity:
\begin{enumerate}
\item
As already mentioned, canonical gravity, whatever its limitations in
other areas, does automatically enforce chronology protection by its
very construction. Canonical quantum gravity certainly has serious
limitations, but it does at least provide a firm kinematic foundation.
\item
Lorentzian lattice quantum gravity, as championed by Ambjorn and Loll,
also enforces chronology protection by
construction~\cite{Ambjorn-Loll,Ambjorn-Loll2,Ambjorn-Loll3,Ambjorn-Loll4}.
It does so by only summing over a subset of Euclidean lattice
geometries, a subset that is compatible with a global Wick rotation
back to globally hyperbolic Lorentzian spacetime. At least in low
dimensionality, large low-curvature regions of spacetime emerge (large
compared to the Planck length, small sub-Planckian curvature). These
regions are suitable arenas for curved-space quantum field
theory. There are however many loose ends to work out --- such as the
details of the emergence of the Einstein--Hilbert action in the
low-energy limit.
\item
Quantum geometry (Ashtekar new variables) is still in a state where
details concerning the emergence of a ``continuum limit'' are far from
settled; in particular it is not yet in a position to say anything
about chronology protection one way or the other.
\item
Brane models (\emph{nee} string theory) are also not yet able to
address this issue. In the low-energy limit brane models are
essentially a special case of semiclassical quantum gravity, with the
brane physics enforcing a particular choice of low-energy quantum
fields on spacetime. In this limit, brane models have nothing
additional to say beyond generic semiclassical gravity. In the
high-energy limit where the physics becomes ``strongly stringy'' the
entire manifold picture seems to lose its relevance, and there is as
yet no reliable formulation of the notion of causality in the string
regime. One possibility is to use string dualities: If the
strongly-coupled string regime is dual to a weakly-coupled regime
where the manifold picture does make sense, then you can at least
begin to formulate local notions of causality in the weakly coupled
regime and then bootstrap them back to the strongly-coupled regime via
duality. But then you still have to decide which class of geometries
you will permit in the weakly-coupled regime (globally hyperbolic?
stably causal?), and the overall situation is far from clear.
\end{enumerate}

So, is chronology protected? Despite a decade's work we do not know
for certain, but I think it fair to say that the bulk of physicists
looking at the issue believe that something along the lines envisaged
by Stephen in his ``chronology protection conjecture'' will ultimately
save the day, as Stephen puts it:
\begin{quotation}
\emph{There is also strong experimental evidence in favour of the
conjecture --- from the fact that we have not been invaded by hordes
of tourists from the future.}
\end{quotation}
It seems to me that approaches based on Novikov's consistency
condition~\cite{Novikov,Novikov2,Novikov3} are now somewhat in
disfavour, largely on philosophical rather than physical grounds. The
same comment applies to attempts at invoking the many-worlds
interpretation of quantum physics, or other ways of radially
re-writing the foundations of physics. Still, despite their relative
unpopularity (or maybe, because of their relative unpopularity) these
more radical alternatives should also be kept in mind as exploration
continues.  Unfortunately, if chronology protection is the answer, we
will have to wander deep into the guts of quantum gravity to know for
certain.



\vspace{20pt}
\hrule width 2in
\vspace{5pt}
\noindent \LaTeXe\ \textsc{cmmp} guide v1.02


\begin{thebibliography}{10}
\expandafter\ifx\csname url\endcsname\relax
  \def\url#1{{\tt #1}}\fi
\expandafter\ifx\csname urlprefix\endcsname\relax\def\urlprefix{URL }\fi
\providecommand{\eprint}[2][]{\url{#2}}

\bibitem{Stephen-CPC}
Hawking, S.W. (1992).
\newblock The chronology protection conjecture.
\newblock {\em Phys. Rev.\/} {\bf D46}, 603--611.

\bibitem{Morris-Thorne}
Morris, M.S. and Thorne, K.S. (1988).
\newblock Wormholes in space-time and their use for interstellar travel: A tool
  for teaching general relativity.
\newblock {\em Am. J. Phys.\/} {\bf 56}, 395--412.

\bibitem{surgery}
Visser, M. (1989).
\newblock Traversable wormholes from surgically modified {Schwarzschild}
  space-times.
\newblock {\em Nucl. Phys.\/} {\bf B328}, 203--212.

\bibitem{examples}
Visser, M. (1989).
\newblock Traversable wormholes: Some simple examples.
\newblock {\em Phys. Rev.\/} {\bf D39}, 3182--3184.

\bibitem{MTY}
Morris, M.S., Thorne, K.S., and Yurtsever, U. (1988).
\newblock Wormholes, time machines, and the weak energy condition.
\newblock {\em Phys. Rev. Lett.\/} {\bf 61}, 1446--1449.

\bibitem{Thorne}
Thorne, K.S. (1992).
\newblock Closed timelike curves.
\newblock In {\em GR13: Proceedings of the 13th Conference on General
  Relativity and Gravitation\/}, eds. R.~J. Gleiser, C.~N. Kozameh, and O.~M.
  Moreschi, pp. 295--315 (Institute of Physics, England).

\bibitem{debate}
Kim, S.W. and Thorne, K.P. (1991).
\newblock Do vacuum fluctuations prevent the creation of closed timelike
  curves?
\newblock {\em Phys. Rev.\/} {\bf D43}, 3929--3947.

\bibitem{Lorentzian}
Visser, M. (1995).
\newblock {\em Lorentzian wormholes: From {Einstein} to {Hawking}\/} (AIP
  Press, USA).

\bibitem{Cosmic-Topology}
Gott, J.R., III (1998).
\newblock Topology and the universe.
\newblock {\em Class. Quant. Grav.\/} {\bf 15}, 2719--2731.

\bibitem{van-Stockum}
van Stockum, W.J. (1937).
\newblock Gravitational field of a distribution of particles rotating about an
  axis of symmetry.
\newblock {\em Proc. R. Soc. Edin.\/} {\bf 57}, 135--154.

\bibitem{Godel}
G{\"o}del, K. (1949).
\newblock An example of a new type of cosmological solutions of {Einstein}'s
  field equations of gravitation.
\newblock {\em Rev. Mod. Phys.\/} {\bf 21}, 447--450.

\bibitem{Tipler-c}
Tipler, F.J. (1974).
\newblock Rotating cylinders and the possibility of global causality violation.
\newblock {\em Phys. Rev.\/} {\bf D9}, 2203--2206.

\bibitem{Hawking-Ellis}
Hawking, S.W. and Ellis, G.F.R. (1973).
\newblock {\em The Large Scale Structure of Space-Time\/} (Cambridge University
  Press, England).

\bibitem{Gott}
Gott, J.R., III (1991).
\newblock Closed timelike curves produced by pairs of moving cosmic strings:
  Exact solutions.
\newblock {\em Phys. Rev. Lett.\/} {\bf 66}, 1126--1129.

\bibitem{Wheeler}
Wheeler, J.A. (1955).
\newblock Geons.
\newblock {\em Phys. Rev.\/} {\bf 97}, 511--536.

\bibitem{Wheeler2}
Wheeler, J.A. (1957).
\newblock On the nature of quantum geometrodynamics.
\newblock {\em Ann. Phys. (NY)\/} {\bf 2}, 604--614.

\bibitem{Geometrodynamics}
Wheeler, J.A. (1962).
\newblock {\em Geometrodynamics\/} (Academic, USA).

\bibitem{Warp}
Alcubierre, M. (1994).
\newblock The warp drive: hyper-fast travel within general relativity.
\newblock {\em Class. Quant. Grav.\/} {\bf 11}, L73--L77.
\newblock \eprint[http://arXiv.org/abs]{gr-qc/0009013}.

\bibitem{Tipler-s}
Tipler, F.J. (1976).
\newblock Causality violation in asymptotically flat spacetime.
\newblock {\em Phys. Rev. Lett.\/} {\bf 37}, 879--882.

\bibitem{Tipler-s2}
Tipler, F.J. (1977).
\newblock Singularities and causality violation.
\newblock {\em Ann. Phys. (NY)\/} {\bf 108}, 1--36.

\bibitem{Tipler-s3}
Tipler, F.J. (1978).
\newblock Energy conditions and spacetime singularities.
\newblock {\em Phys. Rev.\/} {\bf D17}, 2521--2528.

\bibitem{Deser1}
Deser, S., Jackiw, R., and 't~Hooft, G. (1992).
\newblock Physical cosmic strings do not generate closed timelike curves.
\newblock {\em Phys. Rev. Lett.\/} {\bf 68}, 267--269.

\bibitem{Deser2}
Deser, S. and Jackiw, R. (1992).
\newblock Time travel?
\newblock {\em Comments Nucl. Part. Phys.\/} {\bf 20}, 337--354.
\newblock \eprint[http://arXiv.org/abs]{hep-th/9206094}.

\bibitem{Cosmo99}
Visser, M. and Barcel{\'o}, C. (1999).
\newblock Energy conditions and their cosmological implications.
\newblock In {\em Cosmo-99\/}, eds. U.~Cotti, R.~Jeannerot, G.~Senjanovi{\'c},
  and A.~Smirnov, pp. 98--112 (World Scientific).
\newblock \eprint[http://arXiv.org/abs]{gr-qc/0001099}.

\bibitem{Deutsch}
Deutsch, D. (1991).
\newblock Quantum mechanics near closed timelike lines.
\newblock {\em Phys. Rev.\/} {\bf D44}, 3197--3217.

\bibitem{Novikov}
Novikov, I.D. (1992).
\newblock Time machine and selfconsistent evolution in problems with
  selfinteraction.
\newblock {\em Phys. Rev.\/} {\bf D45}, 1989--1994.

\bibitem{Novikov2}
Carlini, A., Frolov, V.P., Mensky, M.B., Novikov, I.D., and Soleng, H.H.
  (1995).
\newblock Time machines: The principle of selfconsistency as a consequence of
  the principle of minimal action.
\newblock {\em Int. J. Mod. Phys.\/} {\bf D4}, 557--580.
\newblock \eprint[http://arXiv.org/abs]{gr-qc/9506087}.

\bibitem{Novikov3}
Carlini, A. and Novikov, I.D. (1996).
\newblock Time machines and the principle of self-consistency as a consequence
  of the principle of stationary action. ii: The cauchy problem for a
  self-interacting relativistic particle.
\newblock {\em Int. J. Mod. Phys.\/} {\bf D5}, 445--480.
\newblock \eprint[http://arXiv.org/abs]{gr-qc/9607063}.

\bibitem{LSH}
{Various~authors},  (1970--1990).
\newblock Tales of the {Legion} of {Super} {Heroes}.
\newblock Marvel Comics.

\bibitem{Zombies}
Heinlein, R.A. (1959).
\newblock All you {Zombies}.
\newblock In {\em The unpleasant profession of {Jonathan} {Hoag}\/} (New
  English Library, USA).

\bibitem{action}
Cassidy, M.J. (1997).
\newblock Divergences in the effective action for acausal spacetimes.
\newblock {\em Class. Quant. Grav.\/} {\bf 14}, 3031--3040.
\newblock \eprint[http://arXiv.org/abs]{gr-qc/9705075}.

\bibitem{Roman-ring}
Visser, M. (1997).
\newblock Traversable wormholes: The {Roman} ring.
\newblock {\em Phys. Rev.\/} {\bf D55}, 5212--5214.
\newblock \eprint[http://arXiv.org/abs]{gr-qc/9702043}.

\bibitem{Grant}
Grant, J.D.E. (1993).
\newblock Cosmic strings and chronology protection.
\newblock {\em Phys. Rev.\/} {\bf D47}, 2388--2394.
\newblock \eprint[http://arXiv.org/abs]{hep-th/9209102}.

\bibitem{Krasnikov}
Krasnikov, S.V. (1996).
\newblock On the quantum stability of the time machine.
\newblock {\em Phys. Rev.\/} {\bf D54}, 7322--7327.
\newblock \eprint[http://arXiv.org/abs]{gr-qc/9508038}.

\bibitem{Automorphic}
Sushkov, S.V. (1997).
\newblock Chronology protection and quantized fields: Complex automorphic
  scalar field in {Misner} space.
\newblock {\em Class. Quant. Grav.\/} {\bf 14}, 523--534.
\newblock \eprint[http://arXiv.org/abs]{gr-qc/9509056}.

\bibitem{Misner}
Tanaka, T. and Hiscock, W.A. (1994).
\newblock Chronology protection and quantized fields: Nonconformal and massive
  scalar fields in {Misner} space.
\newblock {\em Phys. Rev.\/} {\bf D49}, 5240--5245.

\bibitem{KRW}
Kay, B.S., Radzikowski, M.J., and Wald, R.M. (1997).
\newblock Quantum field theory on spacetimes with a compactly generated
  {Cauchy} horizon.
\newblock {\em Commun. Math. Phys.\/} {\bf 183}, 533--556.
\newblock \eprint[http://arXiv.org/abs]{gr-qc/9603012}.

\bibitem{ill-defined}
Cramer, C.R. and Kay, B.S. (1998).
\newblock The thermal and two-particle stress-energy must be ill-defined on the
  2-d {Misner} space chronology horizon.
\newblock {\em Phys. Rev.\/} {\bf D57}, 1052--1056.
\newblock \eprint[http://arXiv.org/abs]{gr-qc/9708028}.

\bibitem{'tHooft1}
Stephens, C.R., 't~Hooft, G., and Whiting, B.F. (1994).
\newblock Black hole evaporation without information loss.
\newblock {\em Class. Quant. Grav.\/} {\bf 11}, 621--648.
\newblock \eprint[http://arXiv.org/abs]{gr-qc/9310006}.

\bibitem{'tHooft2}
't~Hooft, G. (1996).
\newblock The scattering matrix approach for the quantum black hole: An
  overview.
\newblock {\em Int. J. Mod. Phys.\/} {\bf A11}, 4623--4688.
\newblock \eprint[http://arXiv.org/abs]{gr-qc/9607022}.

\bibitem{Reliable}
Visser, M. (1997).
\newblock The reliability horizon for semi-classical quantum gravity: Metric
  fluctuations are often more important than back- reaction.
\newblock {\em Phys. Lett.\/} {\bf B415}, 8--14.
\newblock \eprint[http://arXiv.org/abs]{gr-qc/9702041}.

\bibitem{Reliable2}
Visser, M. (1997).
\newblock The reliability horizon.
\newblock In {\em MG8: Proceedings of the Eighth {Marcel} {Grossmann} Meeting
  on General Relativity\/}, eds. T.~Piran and R.~Ruffini, pp. 608--610.
\newblock \eprint[http://arXiv.org/abs]{gr-qc/9710020}.

\bibitem{Unruh}
Unruh, W. (1981).
\newblock Experimental black hole evaporation?
\newblock {\em Phys. Rev. Lett.\/} {\bf 46}, 1351--1354.

\bibitem{Garay}
Garay, L.J., Anglin, J.R., Cirac, J.I., and Zoller, P. (2000).
\newblock Black holes in {Bose--Einstein} condensates.
\newblock {\em Phys. Rev. Lett.\/} {\bf 85}, 4643.
\newblock \eprint[http://arXiv.org/abs]{gr-qc/0002015}.

\bibitem{Laval}
Barcel{\'o}, C., Liberati, S., and Visser, M. (2001).
\newblock Towards the observation of {Hawking} radiation in {Bose--Einstein}
  condensates. \eprint[http://arXiv.org/abs]{gr-qc/0110036}.

\bibitem{Ambjorn-Loll}
Ambjorn, J. and Loll, R. (1998).
\newblock Non-perturbative {Lorentzian} quantum gravity, causality and topology
  change.
\newblock {\em Nucl. Phys.\/} {\bf B536}, 407--434.
\newblock \eprint[http://arXiv.org/abs]{hep-th/9805108}.

\bibitem{Ambjorn-Loll2}
Ambjorn, J., Correia, J., Kristjansen, C., and Loll, R. (2000).
\newblock On the relation between {Euclidean} and {Lorentzian} 2d quantum
  gravity.
\newblock {\em Phys. Lett.\/} {\bf B475}, 24--32.
\newblock \eprint[http://arXiv.org/abs]{hep-th/9912267}.

\bibitem{Ambjorn-Loll3}
Ambjorn, J., Jurkiewicz, J., and Loll, R. (2000).
\newblock A non-perturbative {Lorentzian} path integral for gravity.
\newblock {\em Phys. Rev. Lett.\/} {\bf 85}, 924--927.
\newblock \eprint[http://arXiv.org/abs]{hep-th/0002050}.

\bibitem{Ambjorn-Loll4}
Ambjorn, J., Jurkiewicz, J., Loll, R., and Vernizzi, G. (2001).
\newblock Lorentzian 3d gravity with wormholes via matrix models.
\newblock {\em JHEP\/} {\bf 09}, 022.
\newblock \eprint[http://arXiv.org/abs]{hep-th/0106082}.

\end{thebibliography}
\end{document}